\documentclass[fleqn,10pt]{wlscirep}
\usepackage[utf8]{inputenc}
\usepackage[T1]{fontenc}
\usepackage{todonotes}
\title{Sharp dose profiles for high precision proton therapy using focused proton beams}

\author[1]{Fardous Reaz}
\author[1 *]{Kyrre Ness Sjobak}
\author[1, 2]{Eirik Malinen}
\author[1]{Nina Frederike Jeppesen Edin}
\author[1]{Erik Adli}

\affil[1]{Department of Physics, University of Oslo, 0316 Oslo, Norway}
\affil[2]{Department of Medical Physics, Oslo University Hospital, P.O. Box 4953 Nydalen, 0424 Oslo, Norway}

\affil[*]{k.n.sjobak@fys.uio.no}

\begin{abstract}
The main objective of radiotherapy is to exploit the curative potential of ionizing radiation while inflicting minimal radiation-induced damage to healthy tissue and sensitive organs.
Proton beam therapy has been developed to irradiate the tumor with higher precision and dose conformity compared to conventional X-ray irradiation.
The dose conformity of this treatment modality may be further improved if narrower proton beams are used.
Still, this is limited by multiple Coulomb scattering of protons through tissue.
The primary aim of this work was to develop techniques to produce narrow proton beams and investigate the resulting dose profiles. 
We introduced and assessed three different proton beam shaping techniques: 1) metal collimators (100/150~MeV), 2) focusing of conventional- (100/150~MeV), and 3) focusing of high-energy (350~MeV, shoot-through) proton beams.
Focusing was governed by the initial value of the Twiss parameter $\alpha$~($\alpha_0$), and can be implemented with magnetic particle accelerator optics.
The dose distributions in water were calculated by Monte Carlo simulations using Geant4, and evaluated by target to surface dose ratio (TSDR) in addition to  the transverse beam size~($\sigma_T$) at the target.
The target was defined as the location of the Bragg peak or the focal point.
The different techniques showed greatly differing dose profiles, where focusing gave pronouncedly higher relative target dose and efficient use of primary protons.
Metal collimators with radii < 2~mm gave low TSDRs (<~0.7) and large $\sigma_T$(>~3.6~mm).
In contrast, a focused beam of conventional (150~MeV) energy produced a very high TSDR (>~80) with similar $\sigma_T$ as a collimated beam.
High-energy focused beams were able to produce TSDRs > 100 and $\sigma_T$ around 1.5~mm.
From this study, it appears very attractive to implement magnetically focused proton beams in radiotherapy of small lesions or tumors in close vicinity to healthy organs at risk.
This can also lead to a paradigm change in spatially fractionated radiotherapy.
Magnetic focusing would facilitate FLASH irradiation due to low losses of primary protons.
\end{abstract}
\begin{document}

\flushbottom
\maketitle

\thispagestyle{empty}

\section*{Introduction}
Radiation therapy~(RT) is one of the most commonly used modalities for curative and palliative cancer treatment~\cite{delaney_role_2005}.
For external beam radiation therapy~(EBRT), high-energy X-rays are most commonly used.
Alternatively, charged particles like electrons, protons, and heavy ions may be used due to their distinct dose profiles.
Protons or heavy ions will deposit a large dose within the last few millimeters of their range.
This peak dose region is known as the Bragg peak~\cite{brown_centenary_2004}.
This feature of heavy charged particles enhances the dose conformity of EBRT compared to X-rays, ensuring better sparing of normal tissue~\cite{patyal_dosimetry_2007}.
Protons are thus  particularly beneficial for treating pediatric cancer, bulky hypoxic tumors, and lesions near the organs at risk (OARs)~\cite{lautenschlaeger_advantage_2019, cotter_proton_2012}.
Using protons for treatment of cancer is not a new concept.
In 1946, Robert R. Wilson first suggested employing high-energy proton beams for treatment~\cite{wilson_radiological_1946}, and the first patient was treated in 1954 at Berkeley Radiation Laboratory~\cite{tobias_pituitary_1958}.
However, proton therapy has gained considerable interest in the last two decades.

Exposure of healthy tissue during EBRT is inevitable.
A range of radiotherapy approaches such as minibeams~\cite{meyer_spatially_2019}, microbeams~\cite{dilmanian_response_2002}, GRID~\cite{henry_proton_2017}, and FLASH~\cite{hughes_flash_2020} have been introduced and studied for the last few years, using proton beams to increase the efficacy of these experimental treatments.
Current proton spot scanning techniques utilize magnetic focusing to shape the beam into a pencil beam of typically 5~mm transverse radius and magnetic sweeping to cover the target.
More advanced beam shaping techniques using an improved beam delivery system are warranted to produce a small spot of radiation dose at a deep-seated target to reduce adverse side effects of EBRT.
However, irradiating a small deep-seated tumor without depositing a significant dose at the surrounding healthy tissue is still challenging due to multiple Coulomb scattering~(MCS), which causes beam broadening and loss of primary protons.
Narrow beams are also an essential element in spatially fractionated radiation therapy (SFRT; also known as GRID), which aims to exploit dose-volume effects in order to reduce normal tissue complications~\cite{billena_current_2019}.

A widely used method for producing narrow proton beams is to use a metal collimator.
However, the dose profiles for narrow collimated beams are degraded transversly due to MCS and subsequent loss of primary protons around the central axis.
This will in turn lead to an unfavorably low target to surface dose ratio (TSDR).
Alternatively, magnetic beam shaping techniques can be used, which reduce the number of secondaries and increase the efficiency of proton delivery~\cite{schneider_advancing_2020}.
Magnetically focused proton beams in the conventional energy range (<~250~MeV; producing a Bragg peak at the target) or with higher energies (>~350~MeV, shoot-through~\cite{verhaegen_considerations_2021} or transmission mode~\cite{van_marlen_ultra-high_2021}) have the potential to reduce the entrance dose, while simultaneously achieving a small spot size at a deep-seated target.
The characteristics of magnetically focused beams of very high energy electrons~(VHEE) have recently been assessed for therapeutic applications~\cite{kokurewicz_focused_2019, whitmore_focused_2021, kokurewicz_experimental_2021}, and high energy focused protons have similar properties but scatter less due to the higher proton mass.

In the work presented here, we have compared three different beam shaping techniques using Monte Carlo~(MC) simulations:
Conventional energy collimated proton~(CECP) beam, conventional energy focused proton~(CEFP) beam, and high energy focused proton~(HEFP) beam.
The overall aim is to demonstrate how sharp focusing can give very favorable dose distributions with the potential to enhance the likelihood of complication-free cure.

\begin{figure}[tbp]
    \centering
    \includegraphics[width=0.95\linewidth]{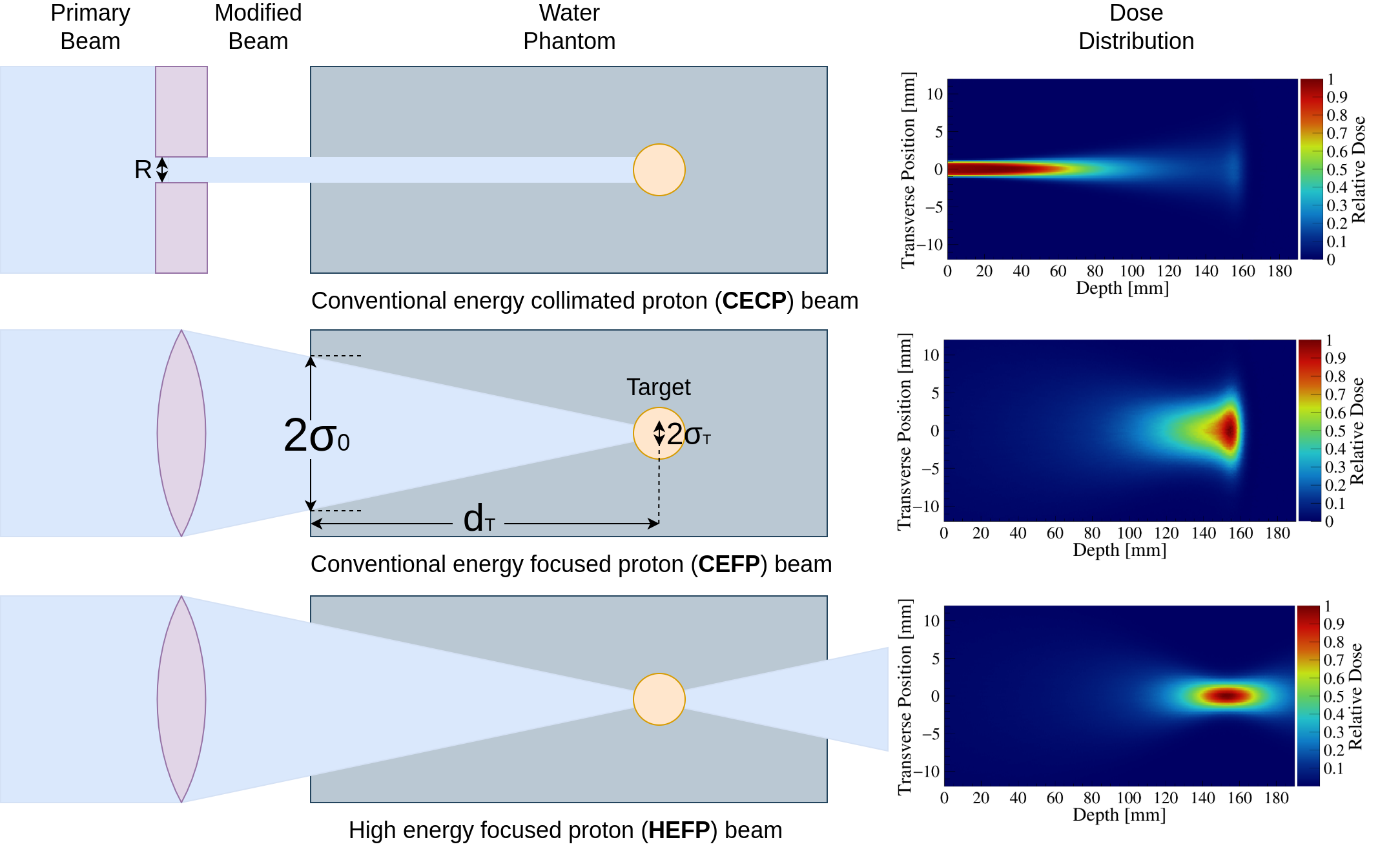}
    \caption{Principle of beam shaping techniques discussed, schematically (left column) and the resulting normalized 2D dose profiles (right column). Also indicated is the meaning of the entrance beam size $\sigma_0$, the target distance $d_T$, the target spot size $\sigma_T$, and the collimator radius $R$.}
    \label{fig:beam_generation}
\end{figure}

The principles of the beam shaping techniques are illustrated in Figure~\ref{fig:beam_generation}.
A metal collimator (top row) can be used to create a narrow beam by eliminating a fraction of the protons delivered from the accelerator.
Alternatively, a group of quadrupole magnets can be used in the beamline to generate focused beams (two bottom rows) with a variable focusing strength, converging the beam so it deposits a large dose at a desired point while the non-specific dose is reduced.
This can be achieved by tuning the magnetic field gradient of the quadrupoles and their relative positions in order to obtain the desired effect on the beam parameters and resulting dose profiles.
Consequently, the proton fluence at the target can be increased, enhancing the TSDR.
For CEFP beams, the intended target is located at the Bragg peak position, which depends on the initial beam energy.
For HEFP beams, the peak dose position does not depend on the beam energy, only on the magnetic focusing parameters, since the Bragg peak is located beyond the patient/phantom geometry.

\section*{Results}

We have studied the properties of CECP and CEFP beams using 100~MeV and 150~MeV initial energies; HEFP beams were simulated with an initial beam energy of 350~MeV.
In all cases, the initial beam is Gaussian and round ($\sigma_x = \sigma_y = \sigma_0$).
For CECP beams, the initial size at the entrance of the collimator was $\sigma_0 = 6.0~\mathrm{mm}$, for CEFP and HEFP the intial size at the entrance of the water phantom was $\sigma_0 = 15~\mathrm{mm}$.
CECP beams were simulated by implementing lead collimators of various inner radii $R$ between 0.5~mm to 6.0~mm.
The focused beams (CEFP \& HEFP) were simulated for different focusing strengths to investigate the impact of beam convergence on the dose profile.
The magnetic focusing system was not explicitly included in the simulations, as it would depend on the beam produced by the accelerator and is thus outside the scope of the study.
The details of the MC simulations and initial beam generation are discussed in the Methods.

\begin{figure}[tbp]
    \centering
    \includegraphics[width=0.31\textwidth]{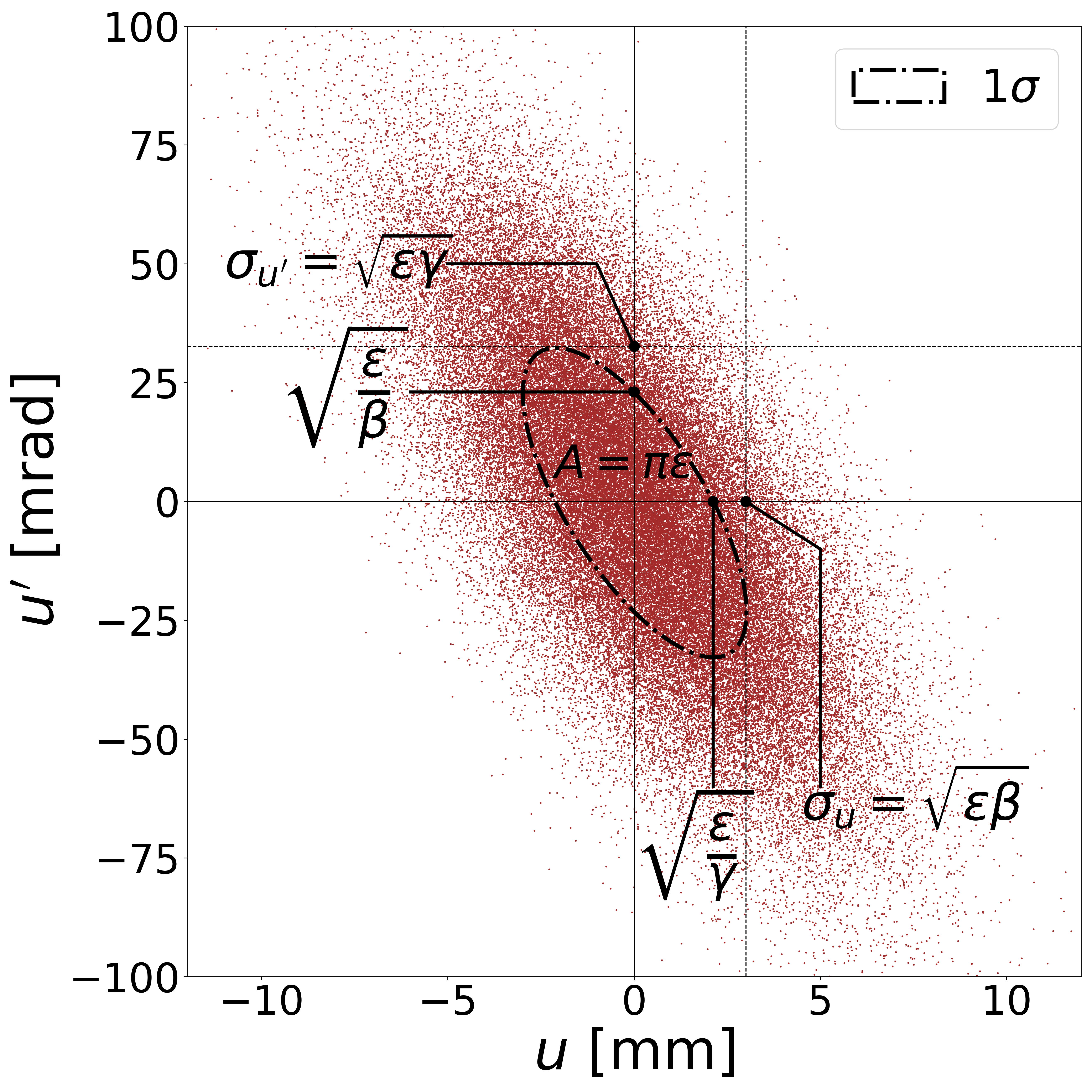}
    \includegraphics[width=0.31\textwidth]{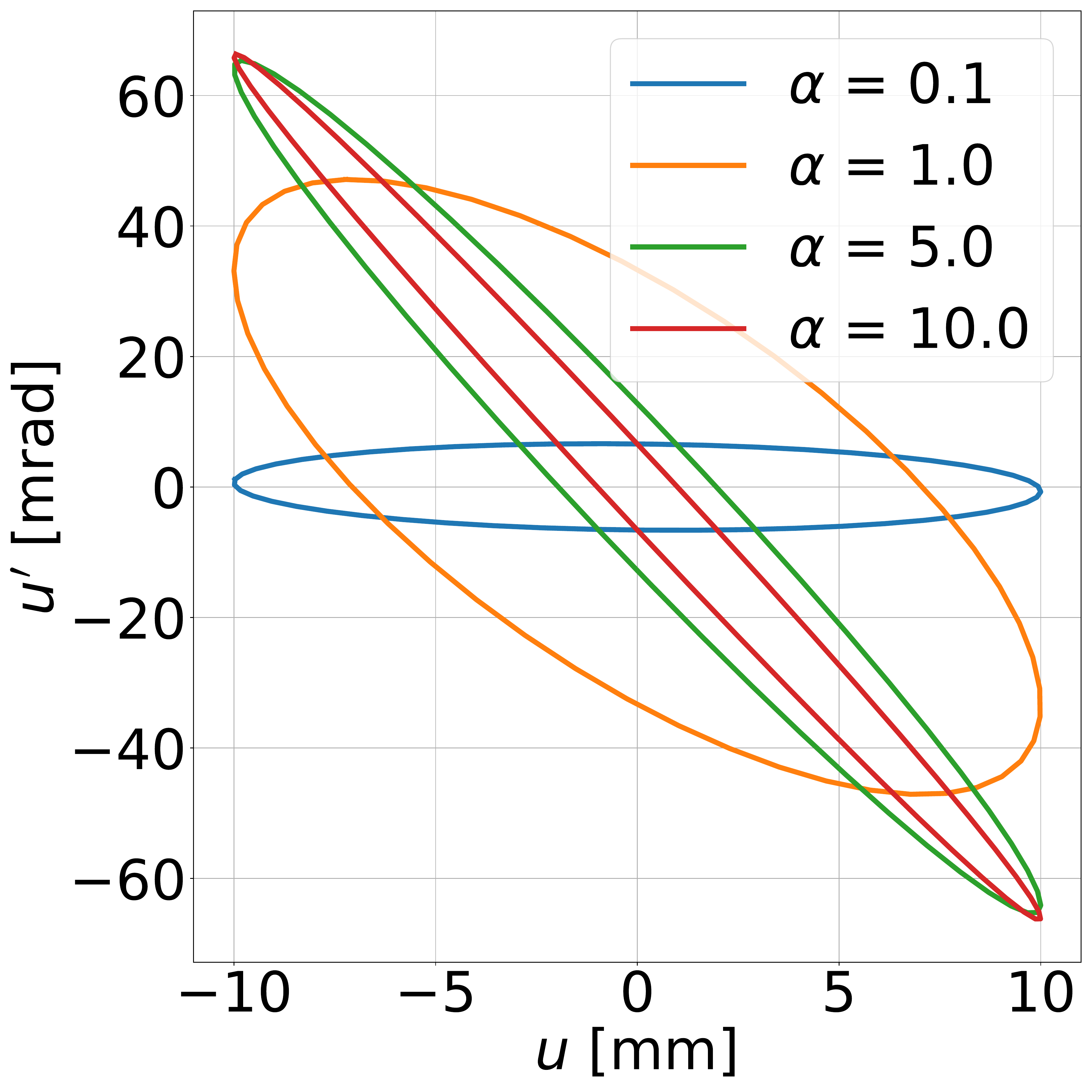}
    \includegraphics[width=0.31\textwidth]{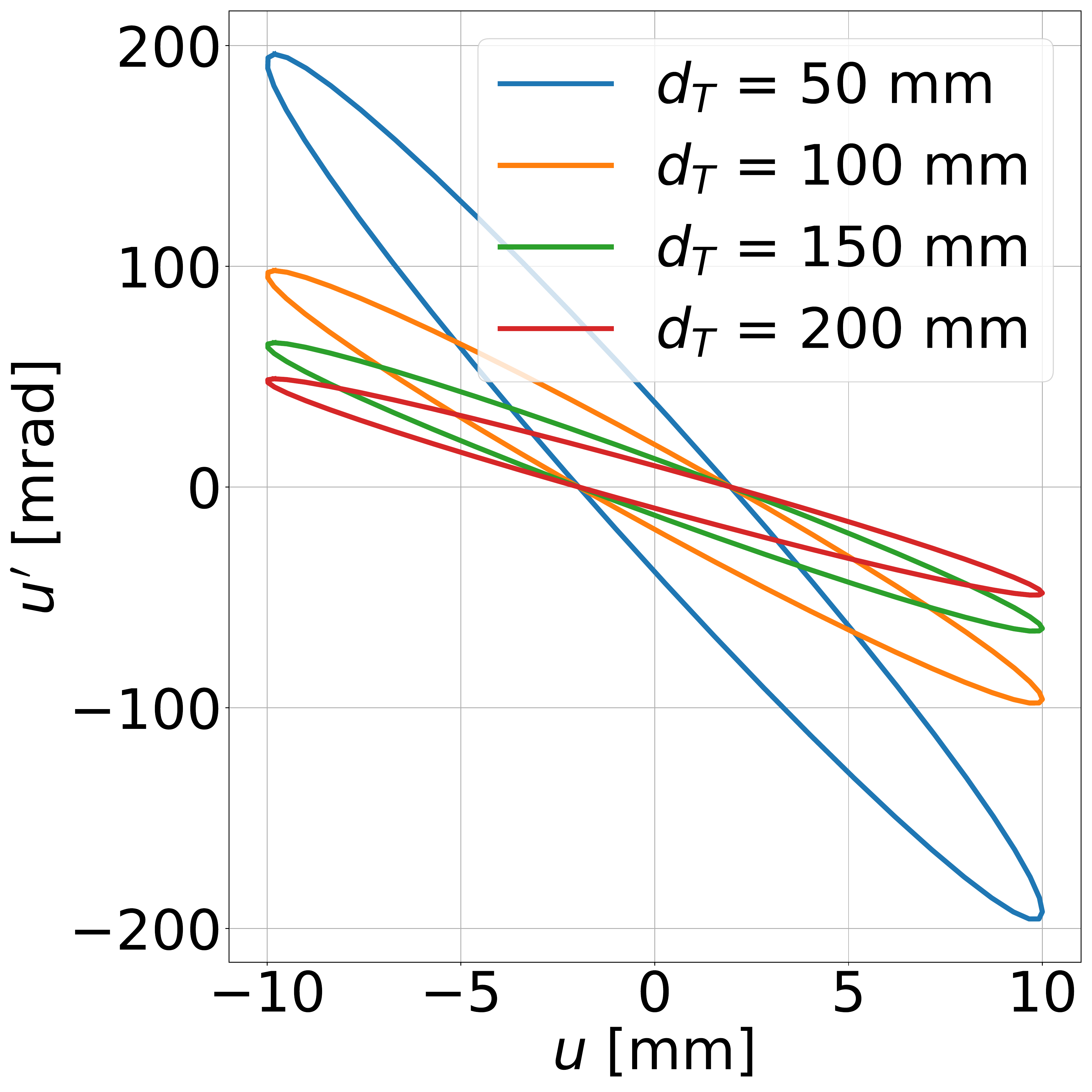}
    \caption{The phase space diagram represents the distribution of the beam's particles. Left:~The Twiss parameters $\alpha$, $\beta$, $\gamma$, and $\epsilon$ describe the Gaussian distribution of the phase-space coordinates $u$ and $u'$ of the particles at a given position along the beam line. The ellipse indicates the $1~\sigma$ boundary of the distribution. Middle:~Phase space ellipse for $\alpha_0$~=~0.1, 1.0, 5.0, and~10.0, at a target distance $d_T = 150~\mathrm{mm}$, and $\sigma_0=10~\mathrm{mm}$. Right:~Phase space ellipses for focused beams with $\alpha_0 = 5.0$ and $\sigma_0=10~\mathrm{mm}$ for target distance 50~mm, 100~mm, 150~mm, and 200~mm.}
    \label{fig:phaseSpace_footprints}
\end{figure}

The phase-space coordinates $u$, $u'$ of the particles are assumed to be distributed with a bivariate Gaussian distribution in both the horizontal and vertical plane; for simplicity we will assume that the distribution in the two planes are identical, giving a round beam.
This distribution can be described using the Twiss parameters~\cite{wille_physics_2001} $\alpha$, $\beta$, $\gamma$, and $\epsilon$.
For the focused beam, the parameter of choice for quantifying the degree of beam focusing is the initial value of $\alpha_0=-\sigma_{u,u'}^2/\varepsilon$.
Here $\sigma_{u,u'}$ is the covariance between the particle position and slope relative to the $z$-axis in the horizontal ($x$) or vertical ($y$) plane.
The variable~$\varepsilon$ is the geometrical root mean square beam emittance, which describes the area of the distribution in phase space.
This is illustrated in Fig.~\ref{fig:phaseSpace_footprints}, showing the initial phase space footprint for several values of $\alpha$ and target point distances $d_t$.
The Twiss parameter $\alpha$ is used to quantify the strength of the focusing, because in vacuum the transverse beam size $\sigma_T$ at the target point only depends on the entrance beam size $\sigma_0$ and the initial $\alpha_0$, and is independent of $d_T$.
The other Twiss parameters are determined by $\sigma_0$, $\alpha_0$, and $d_T$; the details of this is discussed in the Methods.

The transverse magnification of the beam is given as
\begin{equation}
    \frac{\sigma_T}{\sigma_0} = \frac{1}{\sqrt{1 + \alpha_0^2}}\;,
    \label{eq:magnification-alpha}
\end{equation}
where $\sigma_0 > 0$.
For all $\alpha_0 > 0$, the beam at the target is always smaller than the initial beam, so that $\sigma_T / \sigma_0 \le 1$, i.e.\ the beam is de-magnified.

\subsection*{Collimated beams}

\begin{figure}[tbp]
    \centering
    \includegraphics[width=0.9\linewidth]{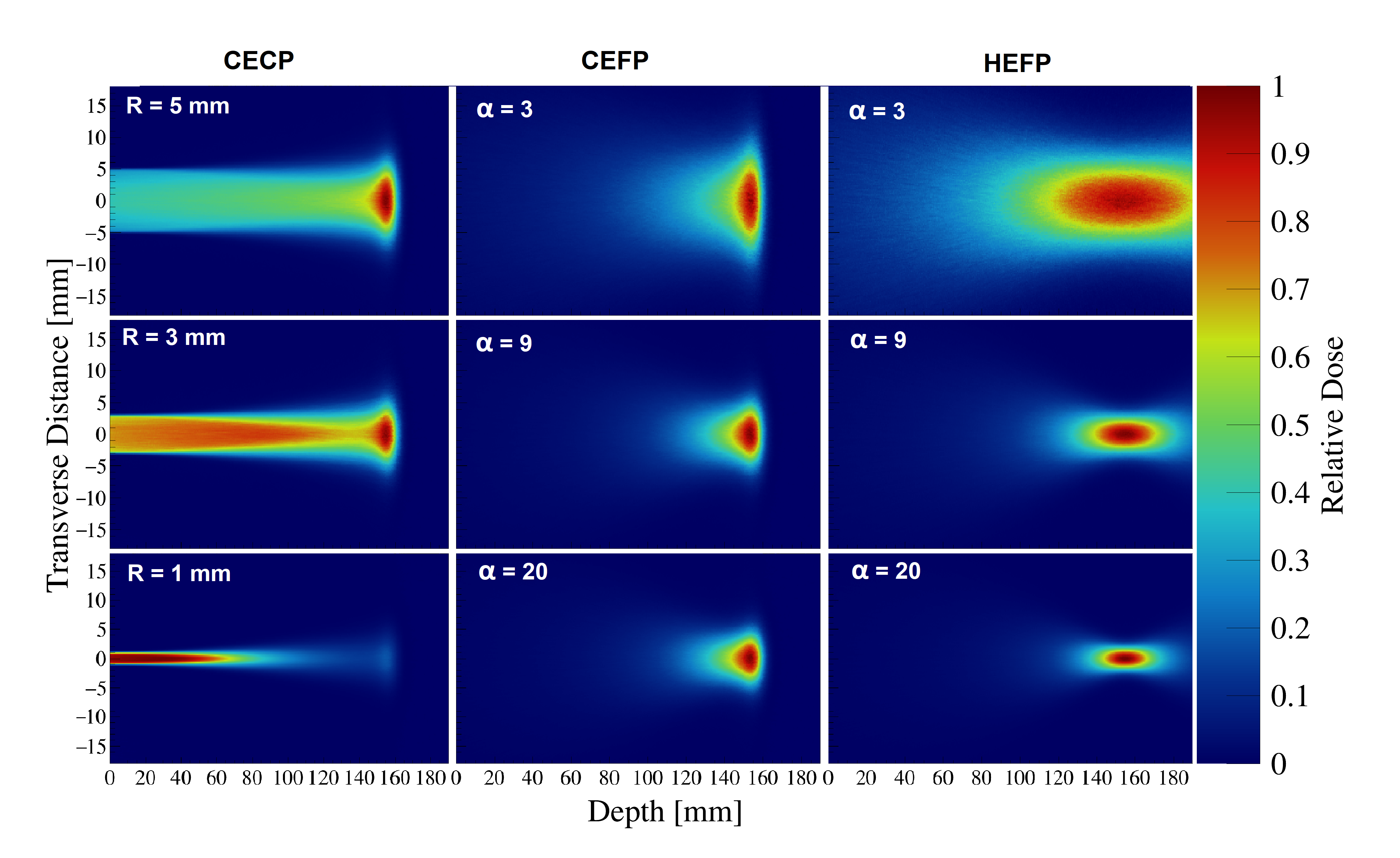}
    \caption{Simulated 2D dose profile for CECP (left column), CEFP (middle) and HEFP (right) beams. CECP and CEFP beam energy is 150~MeV, HEFP 350~MeV. Rows have different beam parameters, either collimator radius $R$ or Twiss parameter $\alpha$.}
    \label{fig:2D_Dose}
\end{figure}
\begin{figure}[tbp]
    \centering
    \includegraphics[width=0.445\textwidth]{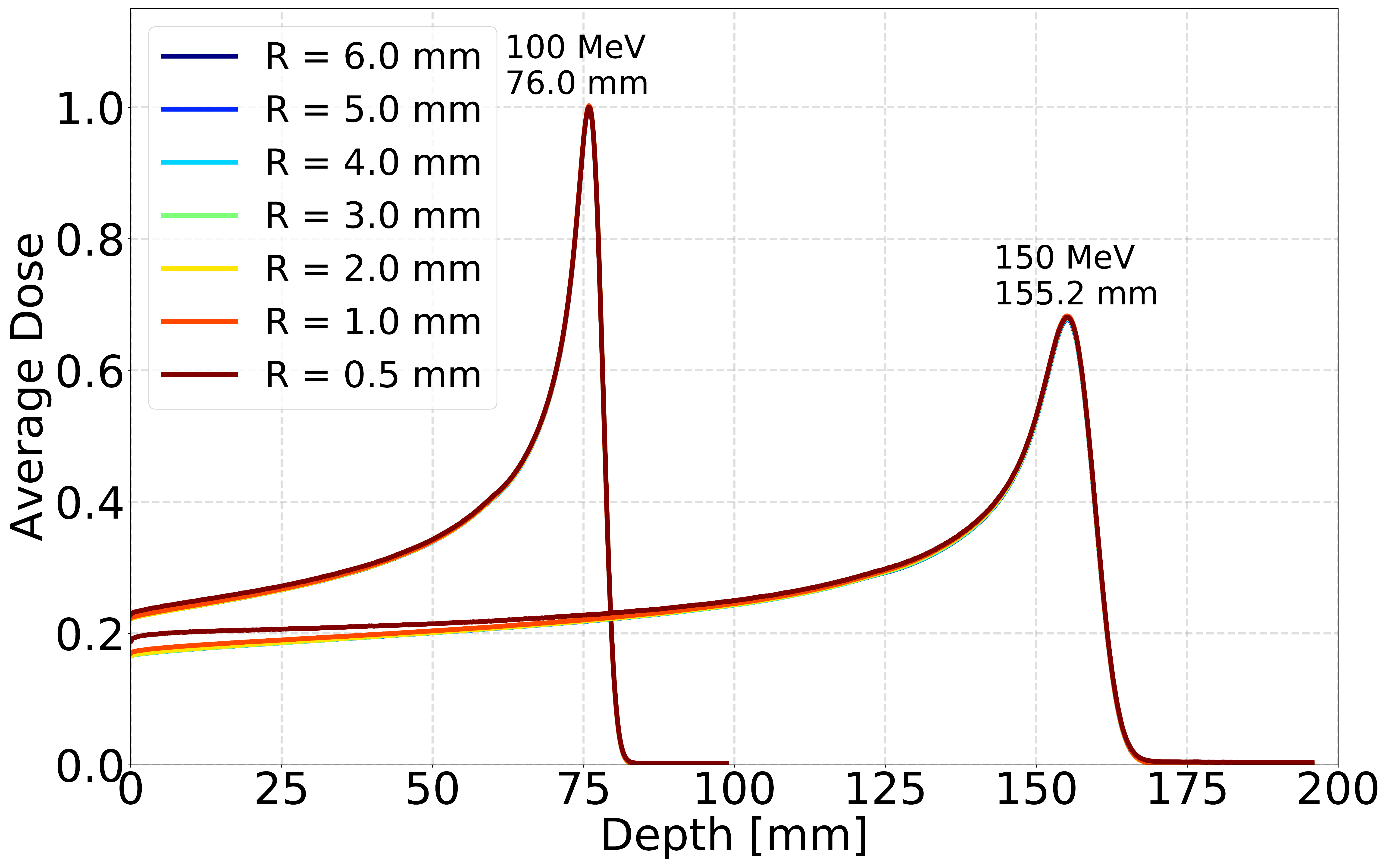}
    \includegraphics[width=0.445\textwidth]{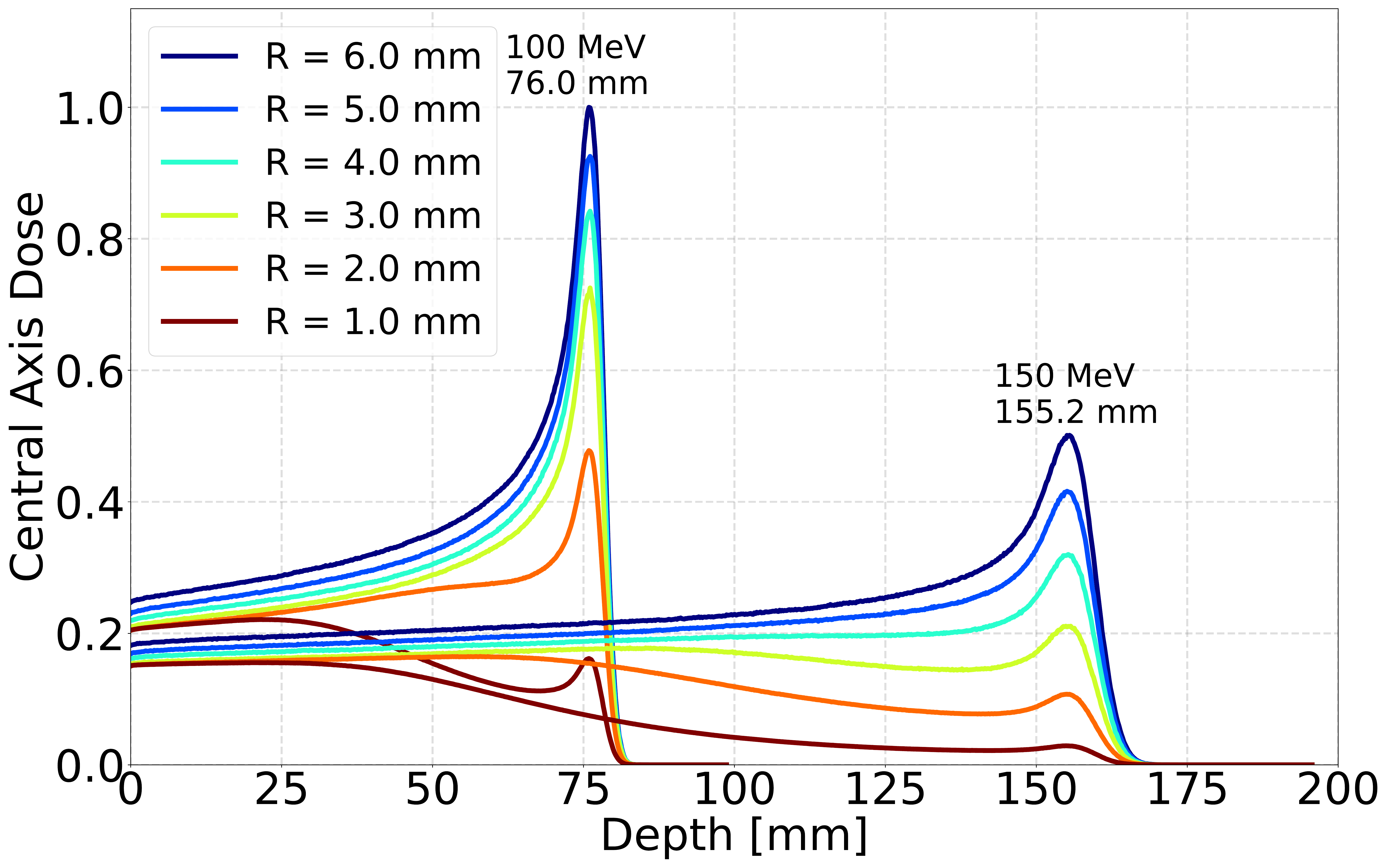}
    \caption{Longitudinal normalized dose profiles for CECP beams. Both wide (left) and narrow (right) cylindrical scoring volumes are used, indicating the overall dose and the central axis dose as a function of depth.}
    \label{fig:Depth-dose_CECP}
\end{figure}
\begin{figure}[tbp]
    \centering
    \includegraphics[width=0.43\textwidth]{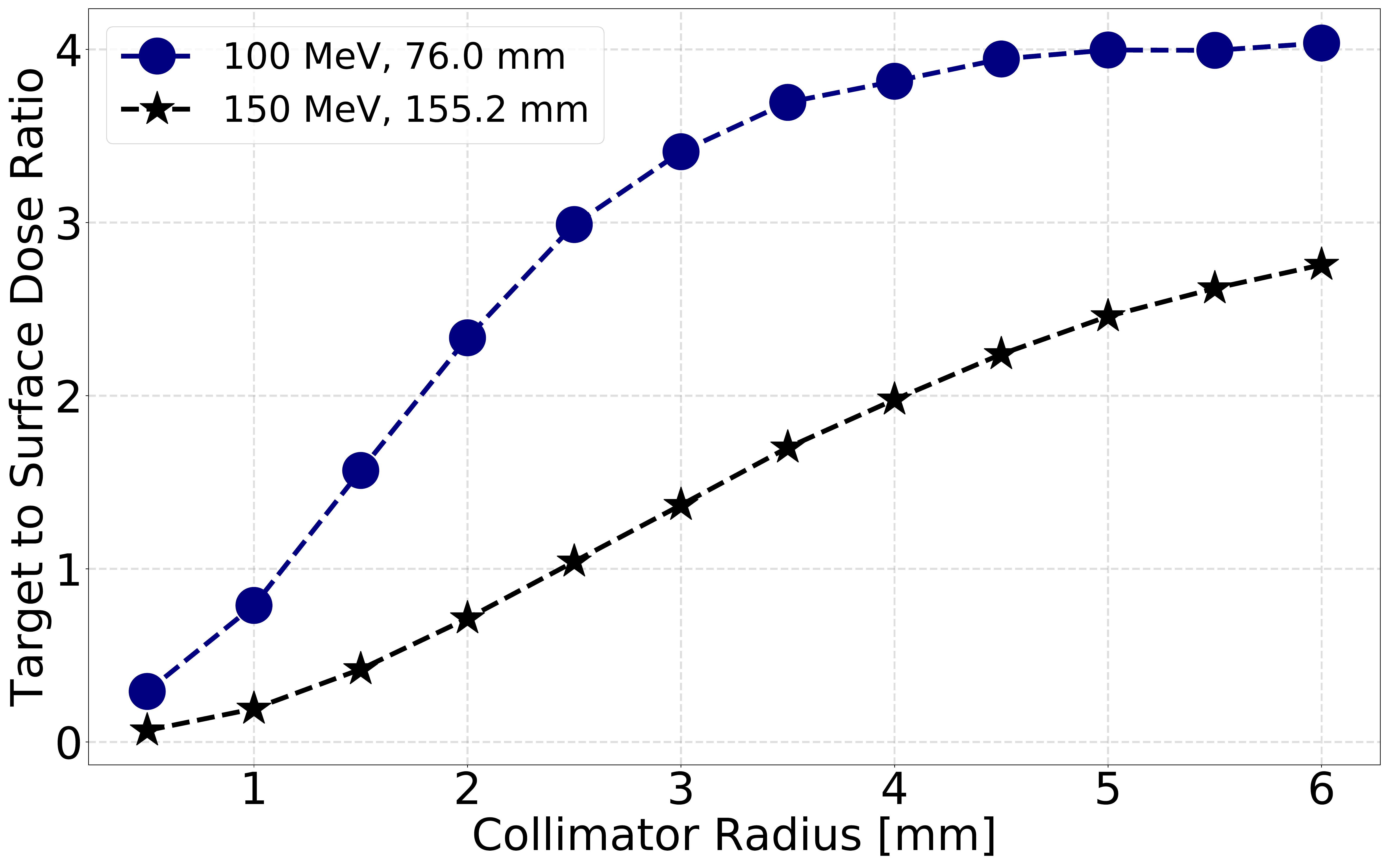}
    \includegraphics[width=0.43\textwidth]{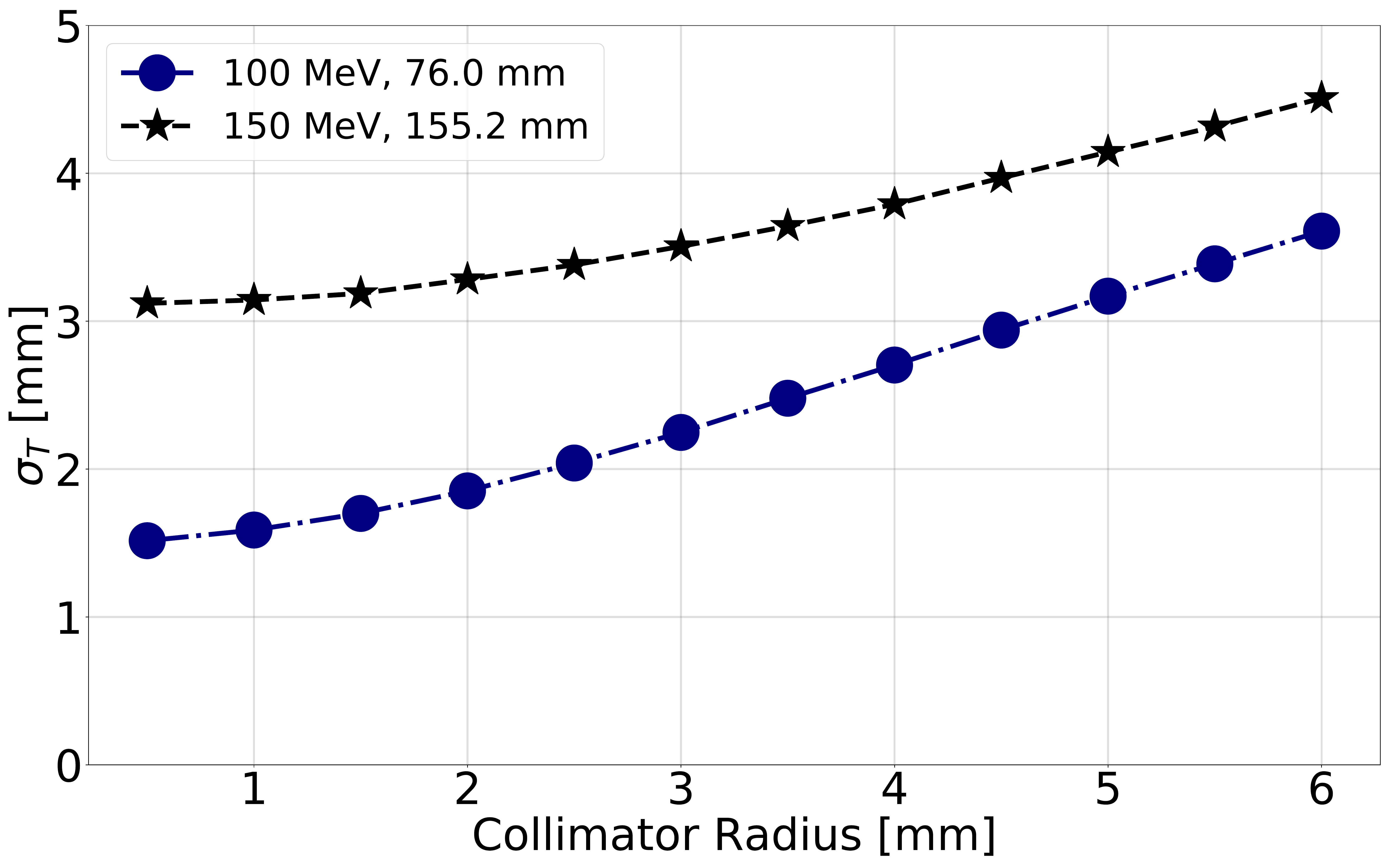}
    \caption{Comparison of 100~MeV and 150~MeV CECP beam: TSDR~(left) and $\sigma_t$ at the Bragg peak position~(right) as a function of collimator radius $R$.}
    \label{fig:FWHM-Size_CECP}
\end{figure}

Figure~\ref{fig:2D_Dose}~(left column) shows the 2D dose profile of CECP beams simulated with an initial energy of 150~MeV, corresponding to target depth of 155.2~mm in water, using collimators with $R=1.0~\mathrm{mm}$, $3.0~\mathrm{mm}$, and $5.0~\mathrm{mm}$.
The details of how the dose profile plots (2D and longitudinal) are generated and normalized is discussed in the Methods section.
A striking difference between the dose distribution of the narrow ($R<3~\mathrm{mm}$) and wide beams is evident.
For narrow beams, the dose at the entrance and in the shallow region are much higher than at the target (Bragg peak), which can be seen in the dose profile of the beam simulated with a 1.0~mm collimator.
This feature is most prominent for the high energy beam, which penetrates deeper and is therefore scattered more.

Longitudinal dose profiles for wide scoring cylinders are depicted in Figure~\ref{fig:Depth-dose_CECP}~(left).
The curves for various collimator radii are almost identical, except for a small discrepancy in dose near the entrance of the water phantom, as the beams stop in the same way.
%The dose contributed by secondary protons generated by the collimator can explain this discrepancy.
As more collimator material is irradiated for a narrow compared to a wide opening,  the fraction of secondary and scattered protons is larger for the narrow collimated beams.
These lower-energy secondary protons are likely to stop within a small depth after the entrance, and cause an elevated dose here.
In addition, the initial beam energy impacts the fractional dose from the secondary protons.
At a higher energy, the dose contribution from primary protons at the entrance is relatively lower because of their smaller stopping power.

Figure~\ref{fig:Depth-dose_CECP}~(right) shows longitudinal dose profiles using a narrow cylindrical scoring volume, displaying how narrowly collimated beams behave very differently compared to the wide beams.
%The dose is normalized to the initial central axis fluence~($F$) of beams.
As seen here, the normalized entrance dose of the narrow beams is higher than its Bragg peak/target dose, which is also apparent from their 2D dose profiles.

The target to surface dose ratio~(TSDR) at the central axis of 100~MeV and 150~MeV CECP beams is shown in Figure~\ref{fig:FWHM-Size_CECP}~(left) as a function of the collimator radius.
The TSDR decreases with decreasing collimator radius, in line with what is visible in Figure~\ref{fig:Depth-dose_CECP}~(right).
In contrast, wider beams deposit a higher dose at the Bragg peak than at the entrance, which explains their larger TSDR.

Figure~\ref{fig:FWHM-Size_CECP}~(right) shows the transverse beam size~($\sigma_T$)  of 100~MeV and 150~MeV CECP beams at the Bragg peak/target for various collimator radii.
This shows that the spot size at the target depends on the depth of the target.
Furthermore, the small initial transverse beam size from a narrow collimator opening is not conserved at the Bragg peak.

\subsection*{Focused beams}

\begin{figure}[tbp]
    \centering
    \includegraphics[width=0.43\textwidth]{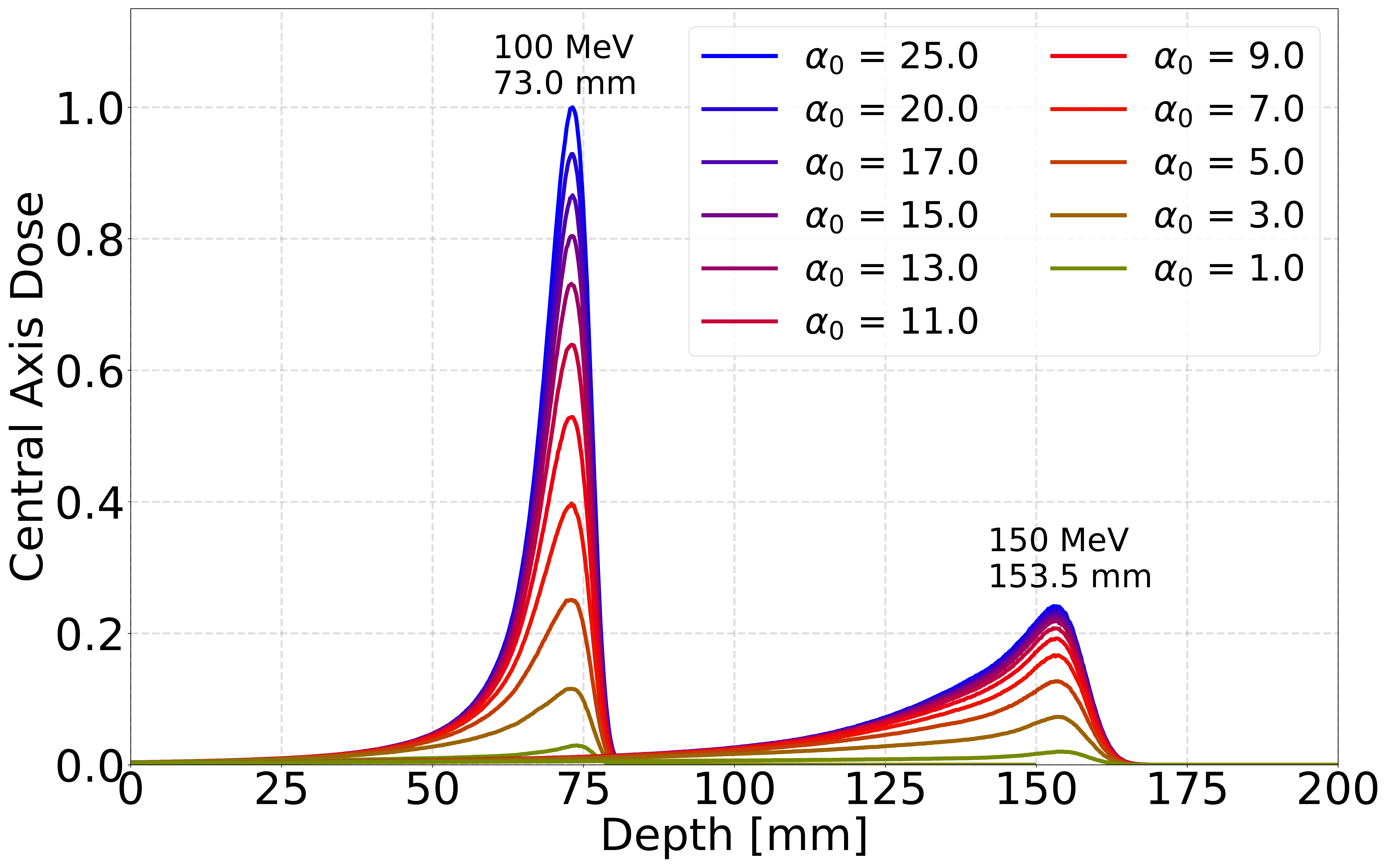}
    \includegraphics[width=0.43\textwidth]{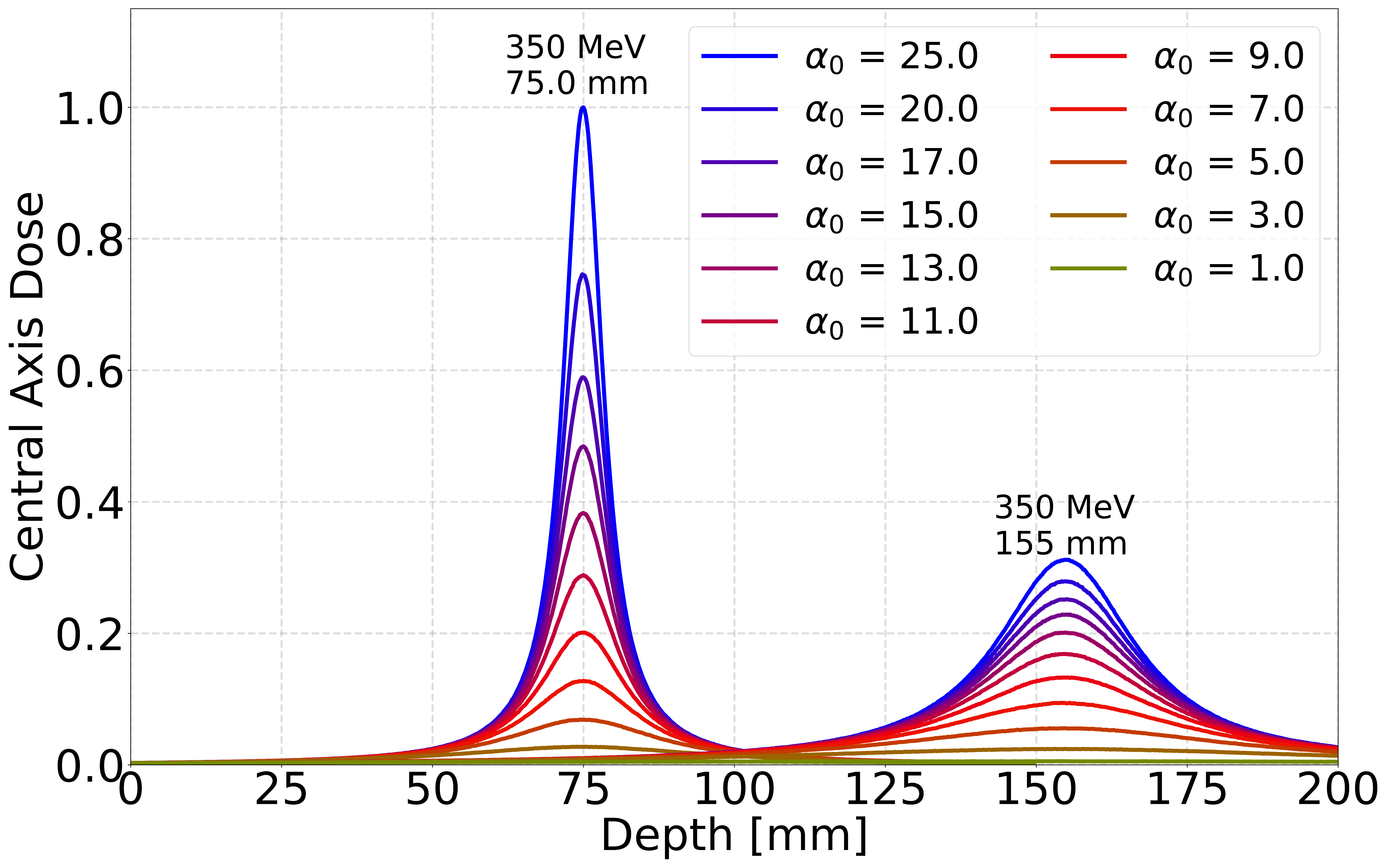}
    \caption{Evolution of central axis dose calculated with a narrow cylinder for CEFP beam~(left) and HEFP beam~(right) as a function of depth.}
    \label{fig:Depth-dose_FB}
\end{figure}
\begin{figure}[tbp]
    \centering
    \includegraphics[width=0.42\textwidth]{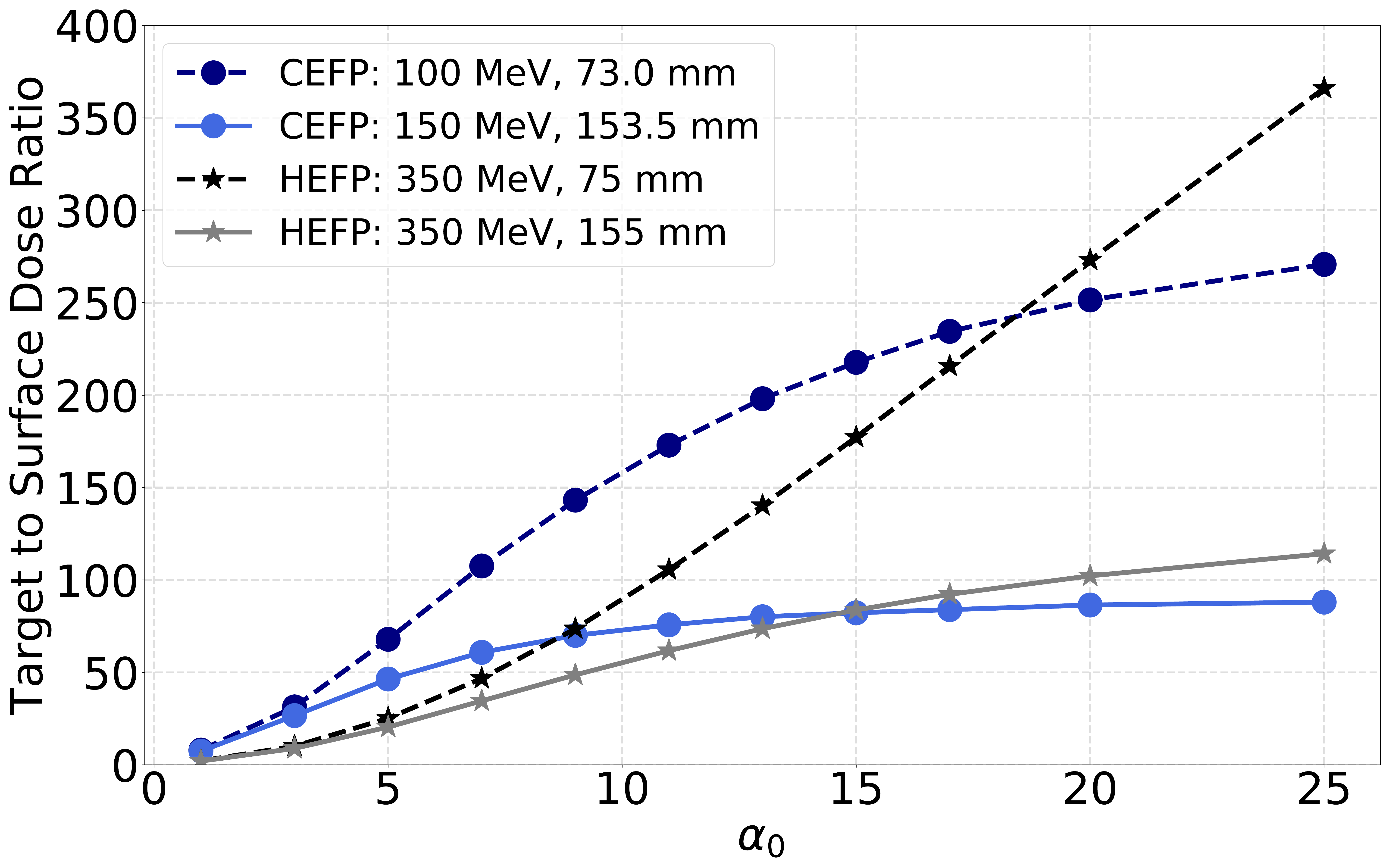}
    \includegraphics[width=0.42\textwidth]{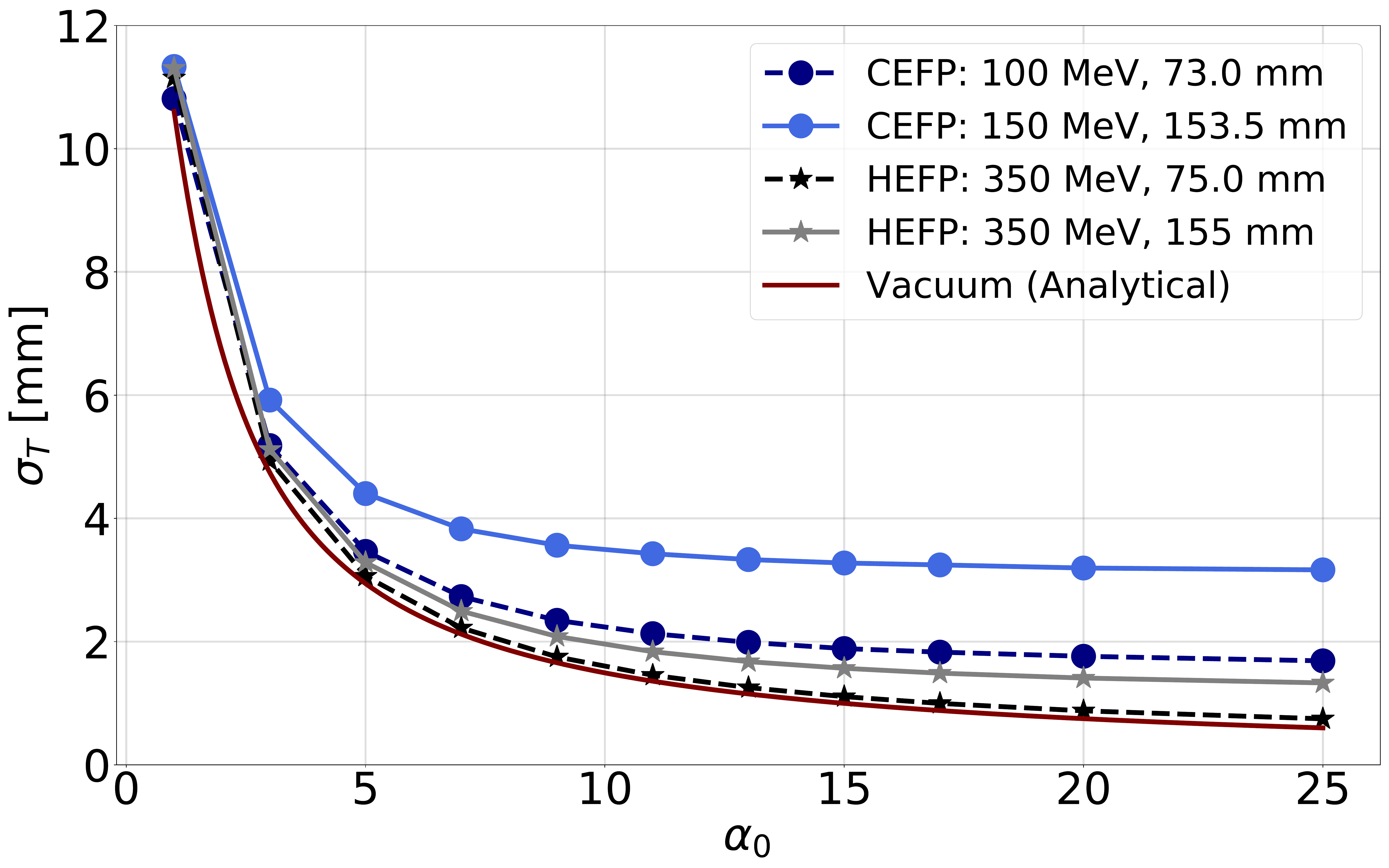}\\
    \caption{TSDR (left) and $\sigma_T$ (right) as a function of focusing strength $\alpha_0$, for two different depths. The beam sizes is also compared with the beam size without scattering expected from Eq.~\eqref{eq:magnification-alpha}.}
    \label{fig:FWHM-Size_FB}
\end{figure}

The 2D dose profile of CEFP beams depicted in Figure~\ref{fig:2D_Dose}~(middle column) is obtained by simulating a 150~MeV proton beam in water with various $\alpha_0$.
Here, the Twiss parameters were set so that the beam is focused at the Bragg peak/target position of the 100~MeV and 150~MeV proton beams, which from the simulation are at 76~mm and 155~mm in water for CECP beams.
In the shallow region where the beam energy is higher, the focusing effect dominates over the scattering effect of MCS, and the beam gradually converges towards the central axis as it propagates in water.
Towards the end of the proton range, the protons are more sensitive to MCS due to the reduced beam energy, causing a significant growth in the transverse beam size relative to what would be expected without scattering.

HEFP beams for a range of $\alpha_0$ were simulated for 350~MeV protons. 
The Bragg peak depth of 350~MeV proton beam is around 665~mm, which is expected to be sufficient to ensure full penetration through most patients.
The other Twiss parameters were set to focus the beam at the same depth as the Bragg peak of a 100~MeV~(76~mm) and a 150~MeV~(155~mm) proton beam, for direct comparison.
The resulting 2D dose profile for a selection of these simulations are shown in Figure~\ref{fig:2D_Dose}~(right column).
Here, the beam gradually converges up to the target, followed by symmetric expansion after the target.
Because the high energy beam is less effected by MCS, a very small spot can be produced at a desired position.
The reduction of the beam energy within the water phantom is not sufficient to create a significant rise in stopping power.
The dose peak is therefore solely due to the concentration of proton fluence near the central axis at the focal point.
Furthermore, in contrast to the CEFP beams, the dose behind the target position is not zero.
Because the total amount of energy deposited per proton is less than the initial energy per proton, the efficiency of the dose delivery is reduced.

The longitudinal depth-dose profiles calculated with a narrow cylindrical scoring volume around the central axis of CEFP and HEFP beams, shown in Figure~\ref{fig:Depth-dose_FB}, have contrasting features in comparison to the CECP beams (Fig.~\ref{fig:Depth-dose_CECP}).
The larger transverse beam size at the entrance reduces the initial fluence of the CEFP and HEFP beams, resulting in a very low relative dose here.
When the beams propagate into the water phantom, the beam size decreases, increasing the fluence near the central axis.
In addition, for CEFP beams (left plot), the energy decreases with depth, increasing the mass-stopping power.
Thus, its higher target dose is a result of both elevated fluence and increased mass-stopping power.
For HEFP beams (right plot), the increased dose at the target is solely due to the increase in fluence around this location, without the presence of the Bragg peak.
For all focused beams, the target dose increases with $\alpha_0$.

The TSDR of 100~MeV and 150~MeV CEFP beams are shown in Figure~\ref{fig:FWHM-Size_FB}~(left).
The TSDR of CEFP beams is enhanced with increasing $\alpha_0$.
Moreover, the ratio achievable for a shallow target (100~MeV) is much larger compared to a deeper target (150~MeV), where the beam is more affected by MCS.
The TSDR of HEFP beams as a function of $\alpha_0$ at 75~mm and 155~mm depth in water are also shown in Figure~\ref{fig:FWHM-Size_FB}~(right).
At both focal points, TSDR increases as a function of initial beam focusing.

In Figure~\ref{fig:FWHM-Size_FB}~(right), the beam size of 100~MeV and 150~MeV CEFP beams decreases with $\alpha_0$, with a smaller size being obtained for the lower energy.
The transverse size of the CEFP beams reaches a minimum plateau at approximately $\alpha_0=11$.
In contrast, the size of HEFP beams at corresponding depths are smaller than for CEFP beams, and can be further reduced by increasing $\alpha_0$.
As seen, HEFP beam with $\alpha_0 \gtrsim 20$ can produce sub-millimetric spots at 75~mm depth.
Thus, HEFP beams can partially overcome the limitation in spot size for deep targets due to reduced scattering at higher energies.
The in-vacuum transverse beam size at the target from Eq.~\eqref{eq:magnification-alpha} is also shown for comparison.
For CEFP beams the deviation is more prominent, due to enhanced MSC near the end of the range of the proton beams.
For HEFP beams, the simulated $\sigma_T$ are very close to the vacuum value, especially for shallow targets.

\subsection*{Comparison}
\begin{table}[tbp]
  \begin{center}
    \caption{Comparison of CECP, CEFP, and HEFP beams at $d_T\approx155~\mathrm{mm}$.}
    \label{tab:compare_modalities}
    \begin{tabular}{lcccc} 
    \hline
    \hline
      \textbf{} & Unit & \textbf{CECP~($R = 2.0~\mathrm{mm}$)} & \textbf{CEFP~($\alpha = 20$)}& \textbf{HEFP~($\alpha = 20$)}\\
      \hline
      Beam energy             & MeV         &   150    &  150    &  350     \\
      Target Depth            & mm          &   155.2  &  153.5  &  155.0   \\
      Surface Dose            & nGy/proton  &   7.16   &  0.063  &  0.037   \\
      Target Dose             & nGy/proton  &   5.095  &  5.362  &  3.703   \\
      TSDR                    &             &   0.71   &  85.7   &  101.0   \\
      FWHM                    & mm          &   7.72   &  7.51   &  3.32    \\
      $\sigma_T$              & mm          &   3.28   &  3.19   &  1.41    \\
      Penumbra~(80\% to 20\%) & mm          &   3.7    &  3.6    &  1.6     \\
      \hline
      \hline
    \end{tabular}
  \end{center}
\end{table}

The properties of three different beam shaping modalities were compared at a depth of $d_T\approx 155$~mm.
Their surface dose and target dose normalized with the number of protons entering the phantom, along with the TSDR, FWHM at target, $\sigma_T$, and transverse penumbra width~(80\% to 20\%) are summarized in Table~\ref{tab:compare_modalities}.

The focused protons will move at an angle relative to the principal horizontal direction, where the initial beam divergence $\sigma_{u'}$ is related to the parameter $\alpha_0$ as
\begin{equation}
    \sigma_{u'} = \frac{\sigma_0}{d_T}\sqrt{\frac{\alpha_0^2}{1 + \alpha_0^2}}\,.
    \label{eq:divergence-alpha}
\end{equation}
The projected depth $d_T$ is therefore slightly smaller than the actual path length.
This leads to a slight difference in the target depth between the CECP and CEFP beams when the beam energy is the same.
For the HEFP beam, the peak dose region is longitudinally elongated at deep-seated targets as seen in Fig.~\ref{fig:2D_Dose}, making it difficult to determine the exact peak dose position.
The transverse beam size and peak dose of CECP and CEFP beams at 155~mm depth were similar, but the surface dose of the CEFP beam was less than 2\% of the CECP beam.
Thus, the TSDR is greatly enhanced for the latter. 
In contrast, the HEFP beam produced the smallest spot with the lowest surface dose, but the target dose per proton is smaller as the beam carries a significant amount of energy out of the phantom.
Still, the TSDR was by far the highest for this beam.

\section*{Discussion}

A metal collimator of appropriate material and dimension can produce narrow CECP beams for proton therapy.
However, physical collimators can not ensure a small spot at a deep-seated target as this is limited by MCS.
For CEFP beams, the minimum achievable target spot size is also limited by MCS.
Still, CEFP beams have a clear advantage of lower entrance dose compared to CECP beams. 
In contrast, HEFP beams are less scattered and can produce a smaller spot at a given depth, with a reduced entrance dose.
However, unlike CECP and CEFP beams which stop at the target, HEFP beams deliver a non-zero dose behind the target.
The TSDR of focused proton beams (CEFP and HEFP) can be enhanced using stronger focusing, giving a higher ratio between the initial beam size and the spot size at the target.
For CECP beams, the TSDR deteriorates as the collimator radius decreases.

In general, as the proton beam propagates into the water phantom, the protons are deflected due to MCS.
The fraction of primary protons scattered away from the central axis of the beam is most significant for the narrowest and deepest CECP beams.
Consequently, the fluence at the center of the beam rapidly decreases with depth.
Simultaneously, the mass-stopping power~($S/\rho$) of the proton beam increases with depth, as a consequence of decreasing energy of the beam.
However, the impact of the drop in fluence is considerably higher than the increase in stopping power.
Therefore, the central axis dose~($D = \Phi \times S/\rho$) of a CECP beam decreases as a function of the depth in water, where $\Phi$ is the proton fluence.
In case of wider beams, the peripheral protons are simultaneously scattered inward, partially balancing or mitigating the outward scattering of protons along the central axis.
Ignoring the primary beam loss by nuclear interactions, an approximate transverse equilibrium can therefore be achieved for wide beams, and the fluence around the central axis of broad beams remains almost constant with depth.
In this case the central axis dose is determined solely by the stopping power, and produces a typical depth-dose curve with a distinctively higher dose at the Bragg peak near the proton beams' range.
Similar dose distributions from narrow collimated proton beams have been observed in several experimental and simulation based microbeam and minibeam studies~\cite{schneider_advancing_2020, girst_proton_2016, lansonneur_first_2020, dilmanian_charged_2015}.

Despite the fact that HEFP beams deposit non-zero doses behind the target, they may have significant advantages over conventional energy proton beams which stop at the target.
The position of the peak dose is in this case set by the focusing magnets, and not from a combination of the initial beam energy and the material composition of the traversed media.
Thus, HEFP beams is more robust against range uncertainties caused by variability in material density in the beam path~\cite{paganetti_range_2012}.
Also, the reduced effect of MCS for high-energy beams also enhances the transverse precision in dose delivery, achieving sharply defined beam spots with a small penumbra.
Furthermore, as was seen in Fig.~\ref{fig:FWHM-Size_FB}~(right) the spot size at the target and the position closely follows the simple analytical in-vacuum estimates, potentially simplifying treatment planning.
Finally, as a 3D spot scan can be realized though variation magnet strengths alone and with constant beam energy, the speed of the proton delivery process can be significantly increased.

On-going research with very high energy electron~(VHEE\cite{kokurewicz_focused_2019, whitmore_focused_2021, kokurewicz_experimental_2021}) focused beam has shown results similar to what we observed for HEFP beams.
However, the smaller mass of electrons compared to protons makes the electrons more sensitive to scattering, which reduces especially the transverse precision in dose delivery.
Because in both cases the particles do not stop at the target point, the high dose region is extended lengthwise compared to CEFB, depending on beam optics.

Narrow CECP beams deliver more secondary protons of low energy due to scattering in the collimator.
As these secondary protons have a short range, the fluence decreases rapidly after the entrance.
However, the stopping power and thus the linear energy transfer (LET) of the lower energy protons is higher compared to the LET of the primary beam.
Thus, dose-averaged LET~($\mathrm{LET}_d$) of narrow beams can be elevated at the entrance because of the secondary protons.
This can potentially give an unwarranted higher relative biological effectiveness (RBE) in this region~\cite{rorvik_exploration_2018}.
The large dose and higher RBE of narrow CECP beams might be detrimental for skin and superficial tissues, and may limit the maximum delivered dose per fraction at the target.
In addition, CECP beams can produce more secondary neutrons than the focused beam due to extensive irradiation of the collimator~\cite{reaz_advanced_2021}.
Despite that the dose contribution of the secondary neutrons is very small compared to the total deposited dose, this might be important for radiation protection due to the high RBE of neutrons with respect to second cancer risk.

A constant RBE of~1.1 is currently employed in state-of-the-art proton therapy. 
However, there is increasing evidence that protons gives a variable RBE that increase with LET, which again increases with decreasing kinetic energy~\cite{rorvik_exploration_2018}.
For HEFP beams, the beam energy is rather high throughout the patient, giving an approximately constant and relatively low LET.
Consequently, a constant RBE can be used for biological dose calculation with higher confidence for this beam configuration.

Use of a collimator to control the beam size in CECP eliminates a fraction of the protons delivered by the accelerator, thus reducing the efficiency of the system.
Consequently, the maximum achievable dose rate decreases. 
An ultra-high dose rate~($>$40~Gy/s) has been recommended to achieve a FLASH effect~\cite{hughes_flash_2020}.
Moreover, a high dose rate can be utilized to overcome the loss of precision caused by tumor motion. 
The CEFB technique can utilize a larger fraction of accelerated protons to deposit a dose at the target compared to CECB, facilitating proton FLASH therapy.
However, for HEFB the ultimate dose rate is reduced since much of the beam energy is deposited outside of the patient.
This is however partially compensated at low $d_T$ since no lossy energy degrading system is needed, increasing the technically achievable beam current.
Even though focused beams can reduce the entrance dose for small targets, the irradiation of large targets with a homogeneous dose requires several beams.
These beams will overlap near the entrance, reducing the TSDR advantage of the focused beams.

\section*{Conclusion}
Using Monte Carlo (MC) simulation, we have assessed three different beam shaping techniques for producing small spots at a deep-seated target: Conventional Energy Collimated Proton (CECP), Conventional Energy Focused Proton (CEFP), and High Energy Focused Proton (HEFP).
The focused beams can be used for irradiating small tumors with a higher dose, and can simultaneously reduce the dose at the surrounding off-target areas.
Therefore, focused beams have the potential to treat a target close to an organ at risk as well as large bulky radiosensitive tumors using a homogeneous or inhomogeneous dose distribution.
CECP beams with small transverse size have the limitation of the large entrance dose and an enlarged spot size at the target caused by MCS.
These limitations can be overcome with the use of conventional or high-energy focused beams.
For irradiation of large targets the advantage of reduced entrance dose for focused beams may be reduced since the beams overlap in this region.

\section*{Methods}
\subsection*{Monte Carlo simulation details}
We used Geant4~\cite{allison_geant4_2006} version~10.07.p02 to simulate the interactions of proton beams with water.
The physics list was constructed using a precompiled reference modular list~(QGSP BIC EMY) based on the recommendation for proton beam therapy simulation~\cite{grevillot_optimization_2010, cirrone_hadrontherapy_2011, winterhalter_evaluation_2020}, which activates G4EMStandardPhysics Option3 that uses the Urban multiple scattering model for simulation of particle interactions. 
The maximum step length was tuned to 0.2~mm to optimize the computation time and simulation accuracy, and the secondary particle production cut was 0.7~mm.
A cylindrical water phantom of radius 200~mm and length 400~mm was implemented in the simulation.
The primary circular Gaussian source for CECP was modeled at 1.0~mm away from a lead collimator.
The outer radius of the collimator was 200~mm, and the inner radius was varied to obtained desired beam.
The length of the collimator was optimized for 250~MeV energy to eliminate the desired fraction of protons out of the collimator aperture.
The world volume was filled with vacuum.

\subsection*{Twiss representation of a beam distribution}
At a given point in a beam line, the particle positions are described in the transverse phase space ($u, u'$), where $u$ is either the horizontal ($x$) or vertical ($y$) particle displacement from the beam axis ($z$), and $u'=\mathrm{d}u/\mathrm{d}z$.

The distribution of particles is commonly described as bivariate Gaussian distribution, with if centered in phase space can be described by a covariance matrix $\Sigma$.
This can be parameterized using Twiss parameters~\cite{wille_physics_2001} $\alpha$, $\beta$, $\epsilon$ and $\gamma=(1+\alpha^2)/\beta$, such that
\begin{equation}
  \Sigma = 
  \begin{bmatrix}
        \langle u^{2}\rangle &  \langle uu'\rangle \\
        \langle u'u\rangle &  \langle u'^{2}\rangle
  \end{bmatrix} = \epsilon
  \begin{bmatrix}
       \beta_{u} &  -\alpha_{u} \\
       -\alpha_{u} &  (1+\alpha^2)/\beta
  \end{bmatrix}\,.
  \label{eq:sigma-matrix}
\end{equation}
Following Liouville’s theorem~\cite{wiedemann_particle_2015}, the phase space area, is conserved as long as the particles are only affected by conservative forces.
The projected geometric root mean square emittance of the beam $\epsilon$ is therefore constant in vacuum and in quadrupole magnets.

\subsection*{Transformation of Twiss parameters}
The Twiss parameters evolve as the beam propagates along the accelerator, and in the case of linear optics this can be found analytically~\cite{wille_physics_2001}.
In free space vacuum, the evolution is simply
\begin{eqnarray}
    \label{Equ:Beta_Space_Theory}
    \beta =  \beta_{0} -2\alpha_{0}s + \gamma_{0}s^2\,, \\
    \label{Equ:Alpha_Space}
    \alpha = \alpha_{0} - \gamma_{0}s\,.
\end{eqnarray}

We can define the beam waist or focal point as a point with the smallest $\beta$ where $\alpha \equiv -(\mathrm{d}\beta/\mathrm{d}s)/2 = 0$.
Using this argument, equation~\ref{Equ:Beta_Space_Theory} leads to the distance of the focal point from the reference point, where Twiss parameters are $\beta_{0}$, and $\alpha_{0}$
\begin{equation}
\label{Equ:focus_depth}
d_{T} = \frac{\alpha_{0}\beta_{0}}{1+\alpha_{0}^2} \,.
\end{equation}
Selecting $d_T$ and $\alpha_0$, this determines $\beta_0$.
In turn $\beta_0$ and $\sigma_0 = \sqrt{\langle u^2 \rangle}$ determines $\epsilon$.

\subsection*{Initial beam generation in Geant4}
The Geant4 particle gun was employed to project particles through the simulated cylindrical water phantom. 
Initial parameters of each particle~(energy, position, and direction of momentum) produced by the gun were assigned to obtain the desired distribution. 
The energy of the particles was assigned from a Gaussian distribution with 1\% spread of the nominal beam energy.

For focused beam, we did not employ magnetic field in the Geant4 simulation.
Instead, the intial Twiss parameters were computed for each focal point as discussed above, and desired focused beams were mimicked following aforementioned method.
For CECP beams, we always started with  a parallel circular Gaussian beam of size $\sigma_u=6.0~\mathrm{mm}$ by employing collimators of required opening with a fixed outer radius of 200~mm.
The emittance of the initial beam was calculated for given set of Twiss parameters~($\alpha=0$, $\beta = 18.8~\mathrm{m}$), and an initial covariance matrix $\Sigma$ was constructed for both transverse planes.

Based on the covariance matrix, the initial transverse phase space of the beam was constructed from a bivariate Gaussian distribution.
A desired beam was produced by random sampling of positions corresponding directions from the transverse phase space distribution.
This method of beam generation was designed by following the open-source Geant4 wrapper MiniScatter~\cite{sjobak_miniscatter_2019}.

\subsection*{Dose computation}
For comparison of different beam shaping techniques, we estimated the dose in a volume of interest $V$ as:
\begin{equation}
    D = \frac {1}{\rho V} \sum_{j=1}^{k} \delta E_j,
    \label{eq:dose}
\end{equation}
where $\delta E_j$ is the $j^\mathrm{th}$ energy deposition for a total of $k$ depositions inside $V$ of mass density $\rho$.
The dose was evaluated both in 2D and along the axis of beam propagation.
The contribution of all primary and secondary protons are considered in the dose comparison; the contribution from other particles are negligible compared to proton dose.

For the 2D dose distributions, the scoring volume was a central slice parallel to the direction of beam propagation, as shown in the Fig.~\ref{fig:2D_Dose}.
This was then divided into cubic voxels of size $0.2^3~\mathrm{mm}^3$ in which Eq.~\ref{eq:dose} was evaluated, giving a 2D profile of the dose.
The 2D dose distribution was also used to compute the transverse size of the dose distribution at the target.
We quantified the transverse beam size $\sigma_T$ using the root mean square of the transverse distribution of the dose.
This was estimated by applying a Gaussian fit function from ROOT~\cite{brun_root_1997} and extracting the $\sigma$; in all cases the fit uncertainty was on the order of a few~$\mathrm{\mu m}$.
The full width at half maximum~(FWHM) was derived from this as $\mathrm{FWHM} = \sigma 2\sqrt{2\ln{2}} \approx 2.355\sigma$, since the Gaussian functions fit the distributions well.
The size of the transverse penumbra was also extracted from the same transverse distribution of the dose.

For the longitudinal dose profile, cylindrical volumes of varying radii were used for the dose calculation, subdivided into discs of length 0.2~mm.
The dose in each disc was calculated using Eq.~\ref{eq:dose}.
Here, because the transverse size of the beam changes with the depth, the radius of the volumes are important.
Both narrow~($r=0.5~\mathrm{mm}$) and wide~($r=200~\mathrm{mm}$) cylinders around the beam axis were used.
Narrow cylinders capture the central axis dose as a function of depth, which is affected by any loss of primary photons from the central axis due to MCS.
The wider cylinders shows the averaged dose throughout the channel, which is mostly affected by the variation in stopping power.
The 2D dose distributions are individually normalized to 1.0 at their maximum in order to demonstrate the relative distribution in each case.
For the longitudinal dose profiles, the curves are normalized by the initial proton fluence at the phantom entrance and then scaled globally so that the highest peak in each plot reaches 1.0.

\section*{Data availability}
The datasets used and/or analysed during the current study available from the corresponding author on reasonable request.

\bibliography{references_needed}

\section*{Acknowledgements}

This work was supported by the Research Council of Norway (NFR Grant No. 255196/F50).

\section*{Author contributions statement}
F.R. wrote and ran the simulations and analysis, and wrote most of the manuscript.
K.N.S. conceived the idea of using strongly converging proton beams for improved precision in analogy to VHEE, and advised on Geant4 simulation, and wrote part of the manuscript.
E.M. and N.F.J.E defined the medical physics features used and edited the manuscript.
E.A. initiated the study, supervised F.R. and edited the manuscript.
All authors reviewed the manuscript.

%\section*{Additional information}

\section*{Competing interests}
The authors declare no competing interests.

\end{document}